\documentclass [twoside,12pt]{article}
\usepackage{amsmath}
\usepackage{array}
\usepackage{graphics}
\usepackage{graphicx}
 \usepackage{epsfig}
\usepackage{epstopdf}
\usepackage{amsmath, amsfonts, amssymb}
\usepackage[displaymath, mathlines]{}
\usepackage{mathtools}
\usepackage{cite}
\begin{document}

\title{\bf Cosmic Transit and Anisotropic Models in f(R,T) Gravity}

\author{
S.K. Sahu\footnote{Department of Mathematics, Utkal University, Bhubaneswar-751004, India,         Email:  jaga1282@gmail.com}, 
S.K.Tripathy\footnote{Department of Physics, Indira Gandhi Institute of Technology, Sarang,
Dhenkanal, Odisha-759146, India,  Email: tripathy\_sunil@rediffmail.com },
P.K. Sahoo\footnote{Department of Mathematics, Birla Institute of
Technology and Science-Pilani, Hyderabad Campus, Hyderabad-500078,
India,  Email:  sahoomaku@rediffmail.com}, A. Nath\footnote{Department of Mathematics, Utkal University, Bhubaneswar-751004, India,  Email:  anasuya \_\ nath@yahoo.com}}

\date{ }

\maketitle

\begin{abstract}
Accelerating cosmological models are  constructed in a modified gravity theory dubbed as
$f(R,T)$ gravity at the backdrop of an 
anisotropic Bianchi type-III universe. $f(R,T)$ is a function of the Ricci scalar $R$ and the trace $T$ of the energy-momentum tensor and it replaces the Ricci scalar in the Einstein-Hilbert action of General Relativity. The models are constructed for two different ways of modification of the Einstein-Hilbert action. Exact solutions of the field equations are obtained by a novel method of integration. We have explored the behaviour of the cosmic transit from an decelerated phase of expansion to an accelerated phase to get the dynamical features of the universe. Within the formalism of the present work, it is found  that, the modification of the Einstein-Hilbert action does not affect the scale factor. However the dynamics of the effective dark energy equation of state is significantly affected.
\end{abstract}



\section{Introduction}

Over a period of hundred years, Einstein's General Relativity(GR) has been successful in explaining a lot of astrophysical as well as cosmological phenomena. Recently, the discovery of gravitational waves by the Laser Interferometer Gravitational-Wave Observatory (LIGO) experiments has  greatly supported the predictions of GR \cite{Abott16, Caste16}. However, GR fails to explain the phenomenon of late time cosmic speed up. This failure has triggered to modify GR in two possible ways: firstly, by incorporating an exotic dark energy form in the matter side of the field equations in GR to provide extra repulsive pressure and secondly, by modifying the geometry part of the field equation (i.e modifying the Einstein-Hilbert action) by replacing a general functional form of the Ricci scalar $R$ and/or by coupling some matter field into it. These modified gravity models have attracted a lot of research attention in the last decade because of their success in explaining many issues in astrophysics and cosmology. So far, our knowledge on dark energy is very poor. Dark energy is believed to dominate the mass-energy budget of the universe, it violates the strong energy conditions and it can cluster at largest accessible scale. Besides the conventional choice of the cosmological constant, a good number of dark energy candidates  such as quintessence, k-essence, phantom field, tachyons, Ghost Dark energy, Ricci Dark energy have been proposed in recent years to handle the issue of late time cosmic acceleration. In some recent works, there also have been attempts to unify the dark energy and dark matter under a single platform of unified dark fluid \cite{Ananda06, Xu12, SKT15}.  Still uncertainty persists in getting the exact nature and origin of dark energy.\\
Modified gravity models do not require dark energy to explain late time cosmic acceleration rather this aspect is taken care of by modifying the geometry. In the present work, we have tried to understand the dynamics of an accelerating universe in the back drop of a modified gravity model. In the modified model, the Einstein-Hilbert action contains a  functional $f(R,T)$ in place of $R$, where $T=g^{ij}T_{ij}$ is trace of the energy-momentum tensor \cite{Harko11}. This theory of gravity has attracted a lot of research interest in recent times \cite{Myrz12, Sharif12, Houndjo12, Alva13, Bar14, Bac14}. Jamil et al. have reconstructed some cosmological models in this theory of gravity using the functional form $f(R,T)=R^2+g(T)$ \cite{Jamil12}. Sharif and Zubair investigated perfect fluid distribution and massless scalar field for Bianchi type-I  universe \cite{Sharif12, Sharif14}. In this modified gravity framework, many authors have studied spatially homogeneous Bianchi type cosmological models\cite{KLM14, BM14, Shamir15, Ahmed14,BM16}. Irregularity energy density factor in  $f(R,T)$  theory that disturbs the stability of homogeneous universe has been studied by Yousaf et al.\cite{Yousaf16}. In another work, they have explored the evolutionary behaviour of compact objects \cite{Yousaf16a}. Recently, Sahoo and his collaborators have extensively investigated different aspects of this modified gravity theory and have reconstructed some $f(R,T)$ cosmological models for anisotropic universes \cite{PKS15, PKS16, PKS16a, PKS16b, GPS16, GPS16a}. Some higher dimensional models have also been constructed in recent times in $f(R,T)$ theory \cite{PKS16, Moraes15}.

In our earlier work \cite{ SKS16}, we have obtained some accelerating models in the framework of $f(R,T)$ theory of gravity for an anisotropic Bianchi type III (BIII) universe. In that paper, we were able to integrate the modified gravity field equations in a novel method with the assumption of a power law functional for a directional scale factor. The present investigation is an extension of the previous work with a linear directional scale factor leading to the generation of a logarithmic scale factor. The universe is believed to have undergone a transition from  a decelerating phase to an accelerating phase. The redshift at which this transition occurred is termed here as the cosmic transit redshift. We have explored the cosmic transit redshift to constrain the model parameters.  The paper is organized as follows. The basic formalism of $f(R,T)$ gravity for a perfect fluid distribution is presented in Sect-2. The exact solutions of the field equations are derived for an anisotropic Bianchi type -III universe in Sect-3. Here, we have considered two different functional forms for $f(R,T)$. The model parameters appearing in the scale factor are constrained from the cosmic transit redshift. The kinematic features of the models are discussed in Sect-4. The dynamics of the anisotropic universe for the two different functional forms of $f(R,T)$ with two possible cosmic transit redshifts is assessed through the effective equation of state parameter in Sect-5. In Sect-6, the conclusions of the present work are presented.

\section{Basic formalism}

The field equations of $f(R,T)$ gravity as has been proposed by Harko et al. \cite{Harko11} and has been widely used by others are derived from the action 
\begin{equation}
S=\frac{1}{16\pi} \int f(R,T)\sqrt{-g} d^4x+\int \mathcal{L}_m \sqrt{-g}d^4x,\label{eq:1}
\end{equation}
where $f(R,T)$ is an arbitrary function of Ricci scalar $R$ and the trace $T$
of the energy-momentum tensor $T_{ij}$ of the matter source. $\mathcal{L}_m$
is the usual matter Lagrangian density. In the present work, we have considered the matter Lagrangian density to be $\mathcal{L}_m=-p$, where $p$ is the pressure of the cosmic fluid. We choose the system of units: $G=c=1$ ($G$ is the Newtonian Gravitational constant and $c$ is the speed of light in vacuum). The $f(R,T)$ gravity field equations are obtained from the action $S$ as

\begin{multline}
f_R(R,T)\biggl(R_{ij}-\frac{1}{3}Rg_{ij}\biggr)+\frac{1}{6}f(R,T)g_{ij} =(8\pi-f_T(R,T))\biggl(T_{ij}-\frac{1}{3}Tg_{ij}\biggr)\\
-f_T(R,T)\biggl(\Theta_{ij}-\frac{1}{3}\Theta g_{ij}\biggr)
+\nabla_i\nabla_jf_R(R,T),\label{eq:2}
\end{multline}
where $\Theta_{ij}= -2T_{ij}-pg_{ij}$ and $\Theta=g^{ij}\Theta_{ij}$. $f_R=\frac{\partial f(R,T)}{\partial R}$ and $f_T=\frac{\partial f(R,T)}{\partial T}$ are the partial differentiation of the respective functional with respect to their arguments. $\Box\equiv\nabla^i\nabla_i$, $\nabla_i$ being the covariant derivative. The stress energy tensor for a cosmic fluid with pressure $p$ and energy density $\rho$ is given by $T_{ij}=(\rho+p)u_iu_j-pg_{ij}$, where $u^i=(0,0,0,1)$ is the four velocity vector in co-moving coordinate system.

It is evident from eq. \eqref{eq:2} that the  physical nature of the matter field decides the behaviour of the field equations of $f(R,T)$ theory of gravity. Therefore different choice of the matter source will lead to different cosmological models in $f(R,T)$ gravity. In other words, one can construct viable cosmological models with different choices of the functional $f(R,T)$. However, Harko et al. \cite{Harko11} have constructed three possible models by considering the functional $f(R,T)$ to be either of $f(R,T)=R+2\chi(T)$, $f(R,T)=\chi_1(R)+\chi_2(T)$ or $f(R,T)=\chi_3(R)+\chi_4(R)\chi_5(T)$ where $\chi(T)$, $\chi_1(R)$, $\chi_2(T)$, $\chi_3(R)$, $\chi_4(R)$  and $\chi_5(T)$ are some arbitrary functions of $R$ and $T$. These functions may be chosen arbitrarily and the obtained results may then be matched with observations concerning late time acceleration or can be reconstructed from some plausible physical basis such as energy conditions and cosmic  thermodynamics. Following our earlier work \cite{SKS16}, we  have considered here two  models: $f(R,T)=R+2\chi(T)$  with $\chi(T)=\lambda T$ and $f(R,T)=\chi_1(R)+\chi_2(T)$ with linear functions $\chi_1(R)=\mu R$ and $\chi_2(T)=\mu T$. 

The field equations for the two specific choices reduce to
\begin{equation}
R_{ij}-\frac{1}{2}g_{ij}R=(8\pi+2\lambda)T_{ij}+\lambda(\rho-p)g_{ij}
\end{equation}
and
\begin{equation}
R_{ij}-\frac{1}{2}g_{ij}R=\biggl(\frac{8\pi+\mu}{\mu}\biggr)T_{ij}+\biggl(\frac{\rho-p}{2}\biggr)g_{ij}.
\end{equation}
It is obvious that, because of the presence of linear functions of $R$ and $T$, the above two specific choices of the functional $f(R,T)$ will behave alike and overlap for the specific choices of the model parameters $\lambda=\frac{1}{2}$ and $\mu=1$. In view of this, we expect similar dynamics of the universe in both the models \cite{SKS16}.

\section{Anisotropic models in the modified gravity framework}

The field equations in the modified gravity theory for the two specific choices of the functional $f(R,T)$ in the backdrop of an anisotropic and spatially homogeneous universe modelled through a Bianchi type III metric 

\begin{equation}
ds^2=dt^2-A^2dx^2-B^2e^{-2mx}dy^2-C^2dz^2
\end{equation}
can be explicitly written as 

\begin{eqnarray}
\frac{\ddot{B}}{B}+\frac{\ddot{C}}{C}+\frac{\dot{B}\dot{C}}{BC} &=& \alpha p-\beta \rho,\\ \label{eq:21}
\frac{\ddot{A}}{A}+\frac{\ddot{C}}{C}+\frac{\dot{A}\dot{C}}{AC} &=& \alpha p-\beta \rho,\\\label{eq:22}
\frac{\ddot{A}}{A}+\frac{\ddot{B}}{B}+\frac{\dot{A}\dot{B}}{AB}-\frac{m^2}{A^2} &=& \alpha p-\beta \rho,\\ \label{eq:23}
\frac{\dot{A}\dot{B}}{AB}+\frac{\dot{B}\dot{C}}{BC}+\frac{\dot{A}\dot{C}}{AC}-\frac{m^2}{A^2} &=& -\alpha\rho+\beta p,\\ \label{eq:24}
\frac{\dot{A}}{A}-\frac{\dot{B}}{B} &=& 0.\label{eq:25}
\end{eqnarray}
Here, the directional scale factors $A, B$ and $C$ are considered as functions of cosmic time $t$ only and the exponent $m$ is an arbitrary positive constant. $\alpha$ and $\beta$ are constants and are decided from the choice of the functional $f(R,T)$. In the above field equations, an overhead dot denotes ordinary derivative with respect to $t$. For the first case with $f(R,T)=R+2\lambda T$ we have $\alpha= 8\pi+3\lambda$ and $\beta=\lambda$ whereas for the second case with the choice $f(R,T)=\mu(R+T)$ we can have $\alpha= \frac{16\pi+3\mu}{2\mu}$ and $\beta=\frac{1}{2}$.  Here we have used the fact that the trace of the energy momentum tensor for our model is $T=\rho-3p$.

Incorporation of eq. \eqref{eq:25} into  the field equations eqs.(6)-(9) yields

\begin{eqnarray}
\frac{\ddot{A}}{A}+\frac{\ddot{C}}{C}+\frac{\dot{A}\dot{C}}{AC} &=& \alpha p-\beta \rho, \label{eq:26}\\ 
2\frac{\ddot{A}}{A}+\biggl(\frac{\dot{A}}{A}\biggr)^2-\biggl(\frac{m}{A}\biggr)^2 &=& \alpha p-\beta \rho, \label{eq:27}\\
\biggl(\frac{\dot{A}}{A}\biggr)^2+2\frac{\dot{A}\dot{C}}{AC}-\biggl(\frac{m}{A}\biggr)^2 &=& -\alpha \rho+\beta p. \label{eq:28}
\end{eqnarray}

From eqs. \eqref{eq:26} and \eqref{eq:27} we get
\begin{equation}
\frac{\ddot{A}}{A}-\frac{\ddot{C}}{C}+\biggl(\frac{\dot{A}}{A}\biggr)^2-\frac{\dot{A}\dot{C}}{AC}-\biggl(\frac{m}{A}\biggr)^2 = 0. \label{eq:14}
\end{equation}
It is obvious that eq. \eqref{eq:14} contains two unknowns $A$ and $C$ and an additional assumed condition is required to get a physically viable cosmological model. In a recent work \cite{SKS16}, a power law form for $C$ was assumed i.e. $C=t^n$, where $n$ is a positive constant i.e. $n>0$. In that work, accelerating cosmological models have been constructed for $n \neq 1$. In the present work, we have explored the interesting case with $n=1$ which provides us a logarithmic scale factor. Following the procedure of integration as adopted in Ref.\cite{SKS16} with the choice of $C=t$, we obtain

\begin{equation}
A^2= m^2t^2lnt .\label{eq40}
\end{equation}

The average scale factor for this model becomes
\begin{equation}
a=\left(m^2t^{3}lnt\right)^{\frac{1}{3}} .\label{eq41}
\end{equation}
Consequently, the deceleration parameter $q=-\frac{a\ddot{a}}{\dot{a}^2}$ is obtained as
\begin{equation}
q=-\frac{(3lnt-2)}{(3lnt+1)^{2}}.
\end{equation}
One should note that, the scale factor has two factors: $t^3$ and $lnt$. One dominates at the early phase and the other dominates at the late phase of cosmic evolution. However, the scale factor vanishes at $t=1$. Before this epoch, the volume scale factor becomes negative. In order to get positive volume factor, we wish to shift the time scale so that the time starts from $t=1$, the point of time when the scale factor and the volume scale factor are zero. The only unknown parameter that affects the volume scale factor is the exponent $m$ in the metric. We have shown the behaviour of the scale factor as function of cosmic time for different representative values of $m$ in Figure 1. It is obvious from the figure that, with the increase in $m$, the slope of the scale factor increases. In general, the deceleration parameter, $q$ does not depend on the value of the exponent $m$. However, in Figure 2, we have shown the deceleration parameter as function of the redshift $z=\frac{1}{a}-1$ for two values of $m$. Here we have assumed the scale factor at the present epoch to be $1$. It is interesting to note that, the deceleration parameter evolves with cosmic time from early deceleration to late time acceleration. The values of $m$ have been constrained from the behaviour of $q$ as a function of redshift so that, the deceleration to acceleration transition occurs at either at $z_{da} \sim 0.5$  or at $z_{da} \sim 0.8$. These two values of transition redshift have been chosen from recent predictions from a host of observational data \cite{Mor16, Busca13}. Corresponding to these values of transition redshift, $m$ is constrained to assume the values $0.245$ and $0.185$ respectively. This has been clearly reflected in fig.2. The interesting feature of the model is that, the choice of the functional $f(R,T)$ does not affect the scale factor and the deceleration parameter. While the scale factor depends on the exponent $m$, the deceleration parameter is independent of $m$. One can also infer that, we may get similar scale factor in GR in the backdrop of an anisotropic BIII universe. However, a modified Einstein-Hilbert action affects the dynamics of the universe that can be decided through the effective dark energy equation of state.

\begin{figure}[t]
    \begin{center}
      \includegraphics[width=1\textwidth]{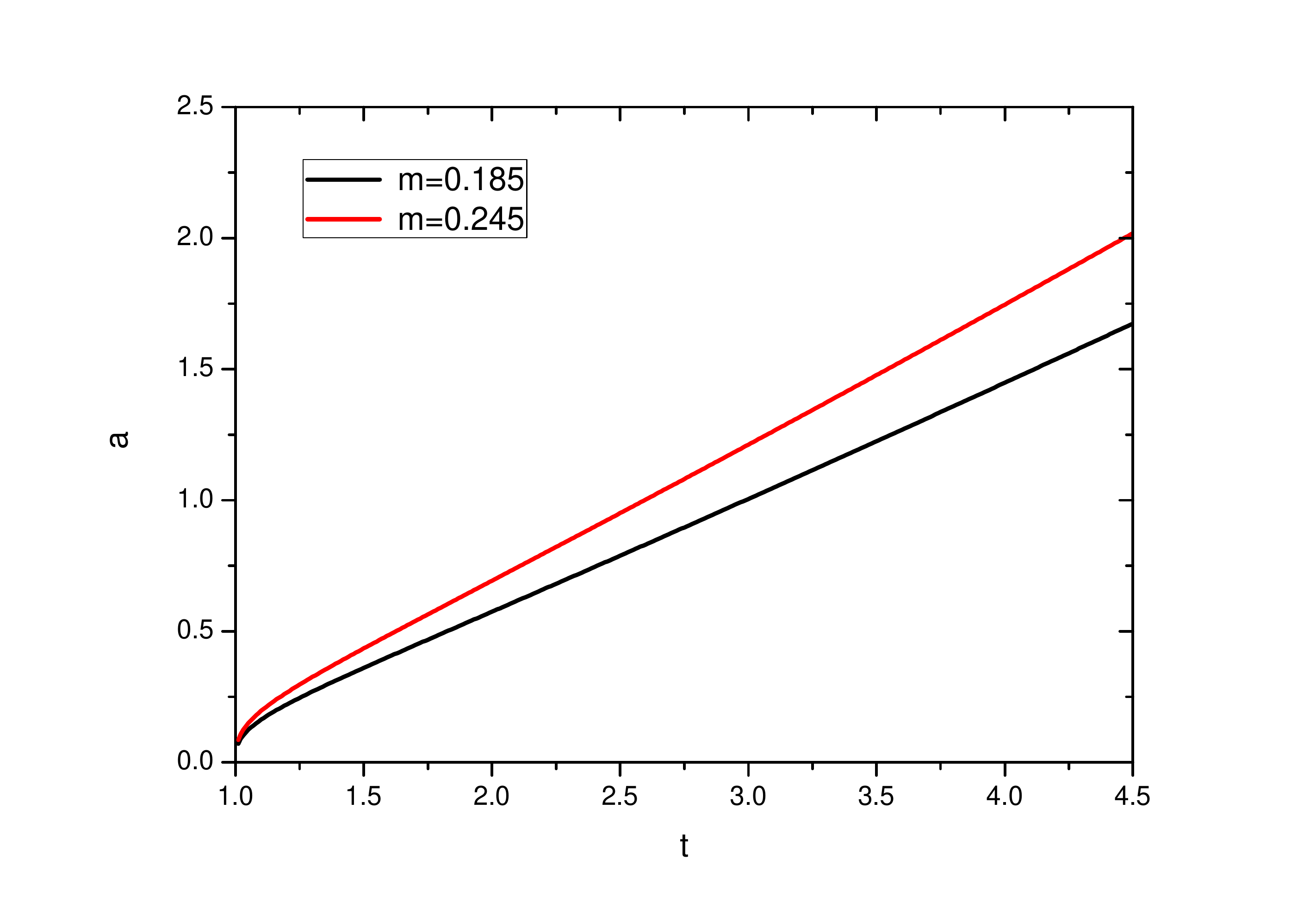}
\caption{Radius scale factor as a function of cosmic time. Radius scale factor depends on the parameter $m$ which has been constrained from two different values of cosmic transit redshift.}
    \end{center}
 \end{figure}

\begin{figure}[h!]
    \begin{center}
      \includegraphics[width=1\textwidth]{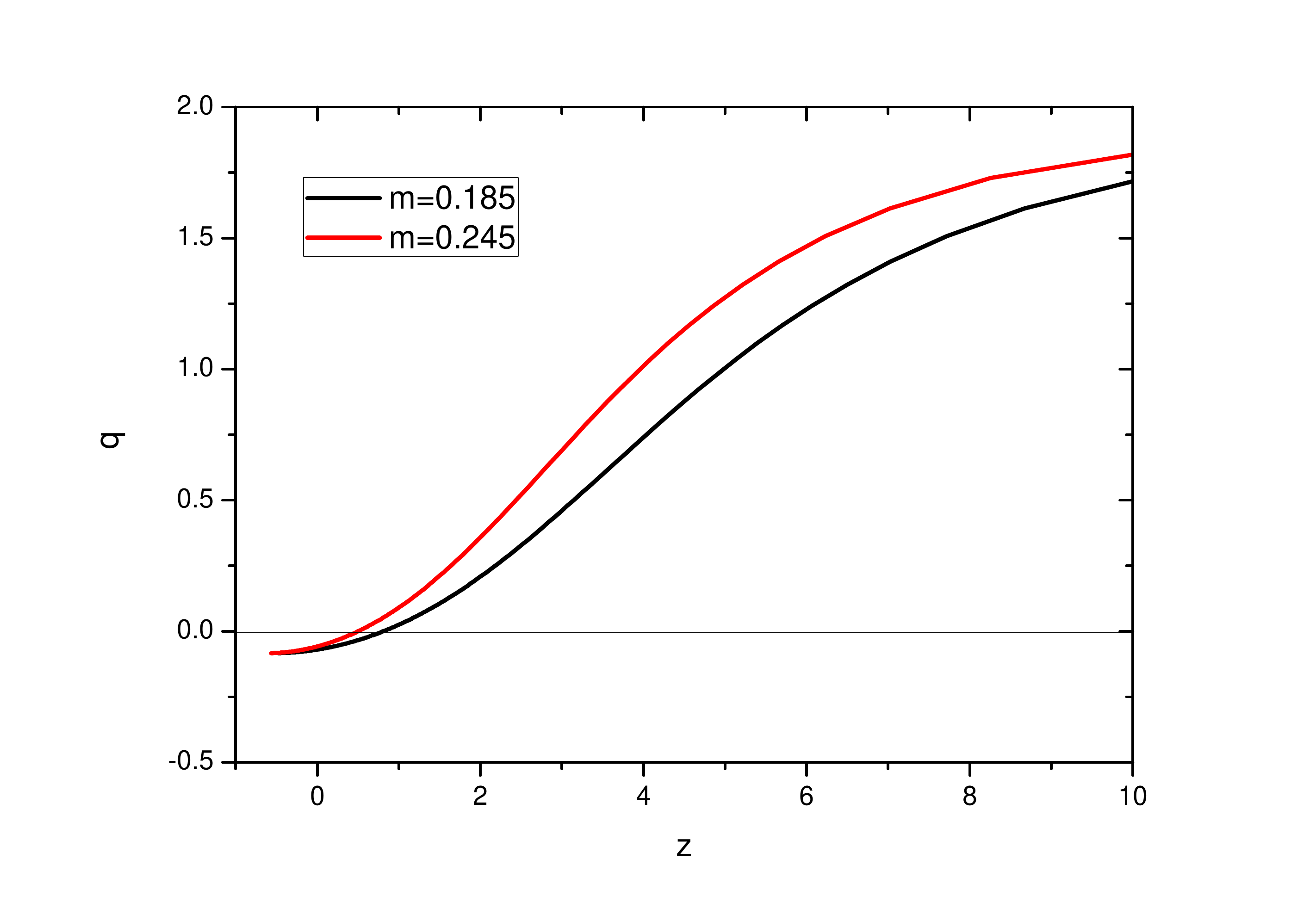}
\caption{Deceleration parameter as a function of redshift for two different values of the parameter $m$.}
    \end{center}
 \end{figure}

\section{Kinematic features}

The scale factor of the model depends on the choice of the exponent $m$ and does not depend on the choice of the functional $f(R,T)$. Consequently the kinematic features of the model for both the choices of the functional $f(R,T)$, as has been considered in the present work, are the same.

In general, for the anisotropic Bianchi -III metric, the scalar expansion $\theta$ and shear scalar $\sigma$ are defined as
\begin{equation}
\theta=u^i_{;i}=\frac{\dot{A}}{A}+\frac{\dot{B}}{B}+\frac{\dot{C}}{C}
\end{equation}
and
\begin{equation}
\sigma^2=\frac{1}{2}\sigma_{ij}\sigma^{ij}=\frac{1}{3}\biggl[\biggl(\frac{\dot{A}}{A}\biggr)^2+\biggl(\frac{\dot{B}}{B}\biggr)^2+\biggl(\frac{\dot{C}}{C}\biggr)^2-\frac{\dot{A}\dot{B}}{AB}-\frac{\dot{B}\dot{C}}{BC}-\frac{\dot{C}\dot{A}}{CA}\biggr].
\end{equation}

The mean Hubble parameter $H$ is
\begin{equation}
H=\frac{1}{3}\biggl(\frac{\dot{A}}{A}+\frac{\dot{B}}{B}+\frac{\dot{C}}{C}\biggr)
\end{equation}
where $H_1=\frac{\dot{A}}{A}, H_2=\frac{\dot{B}}{B}$ and $H_3=\frac{\dot{C}}{C}$ are the directional Hubble parameters in the spatial directions $x,y$ and $z$ respectively.\\
The mean anisotropy parameter $A_m$ is defined as
\begin{equation}
A_m=\frac{1}{3}\sum_{i=1}^3\biggl(\frac{H_i-H}{H}\biggr)^2.
\end{equation}

The kinematic features of the anisotropic model can be well assessed once we obtain the radius scale factor and the consequent directional scale factors. In the present work, we have adopted a very specific method to integrate the field equations to obtain the radius scale factor as $a=\left(m^2t^{3}lnt\right)^{\frac{1}{3}}$. The directional Hubble rates $H_i (i=1,2,3)$ are obtained as 

\begin{equation}
H_1=H_2=\frac{2~lnt+1}{2t~lnt}, \ \ \ H_3=\frac{1}{t},
\end{equation}
so that the mean Hubble parameter $H$ becomes 
\begin{equation}
H=\frac{3~lnt+1}{3t~lnt}.  
\end{equation}

The average anisotropy parameter $A_m$ can be expressed as
\begin{equation}
A_m= \frac{1}{3(3~lnt+1)^{2}}.
\end{equation}
The scalar expansion $\theta$ and the shear scalar $\sigma$ for the model are given by
\begin{equation}
\theta=\frac{3~lnt+1}{t~lnt}, \ \ \ \ \sigma=\frac{1}{2\sqrt{3}t~lnt}.
\end{equation}

The ratio $\frac{\sigma}{\theta}=\frac{1}{2\sqrt{3}(3~lnt+1)}$ goes to zero for large cosmic time implying that the model isotropizes at late times. This feature can also be inferred from the average anisotropic parameter $A_m$ which vanishes for large values of $t$.

\section{Effective dark energy equation of state}

\begin{figure}[h!]
    \begin{center}
      \includegraphics[width=1\textwidth]{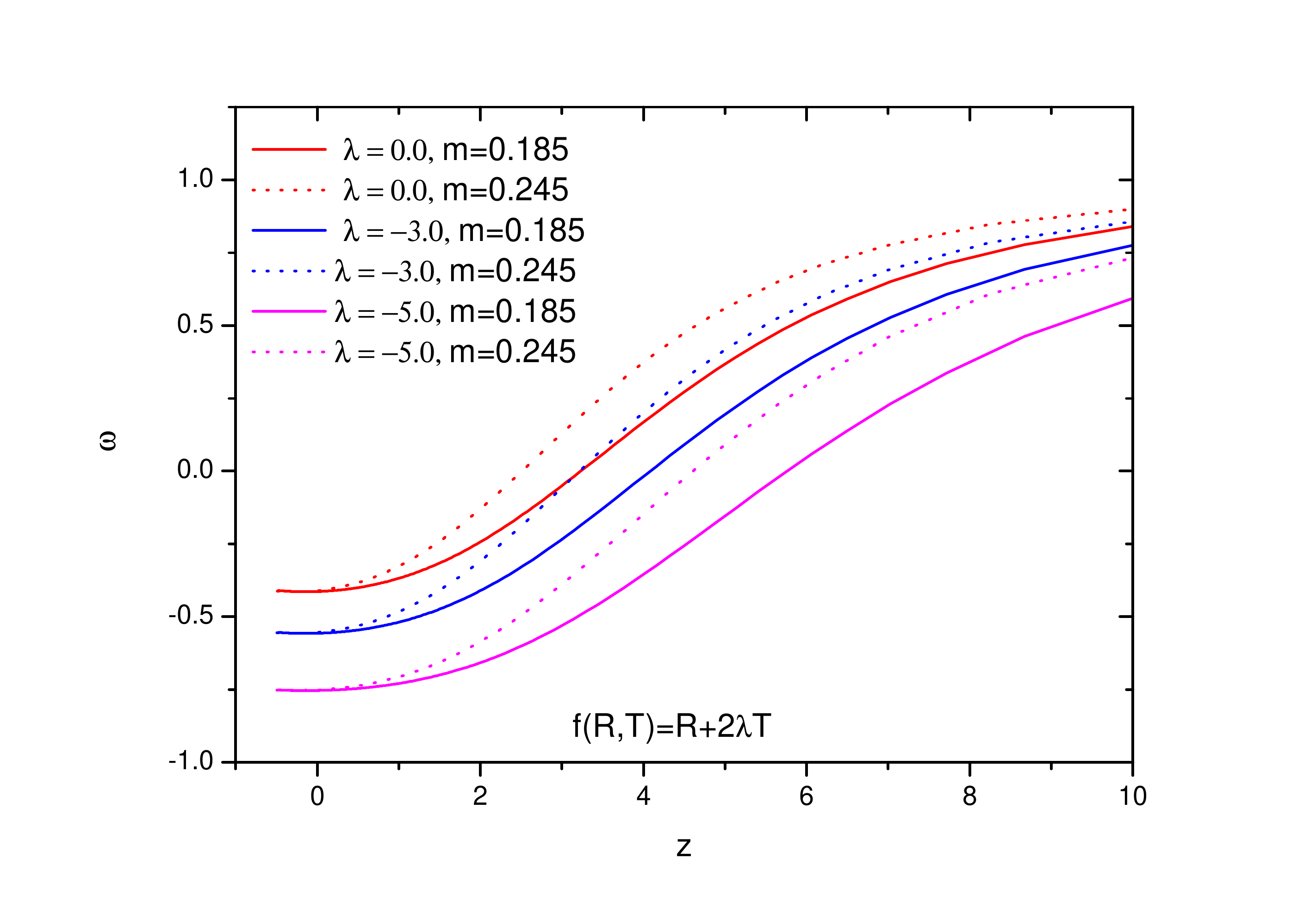}
\caption{Effective dark energy equation of state as a function of redshift for the choice of the functional $f(R,T)=R+2\lambda T$.}
    \end{center}
 \end{figure}

\begin{figure}[h!]
    \begin{center}
      \includegraphics[width=1\textwidth]{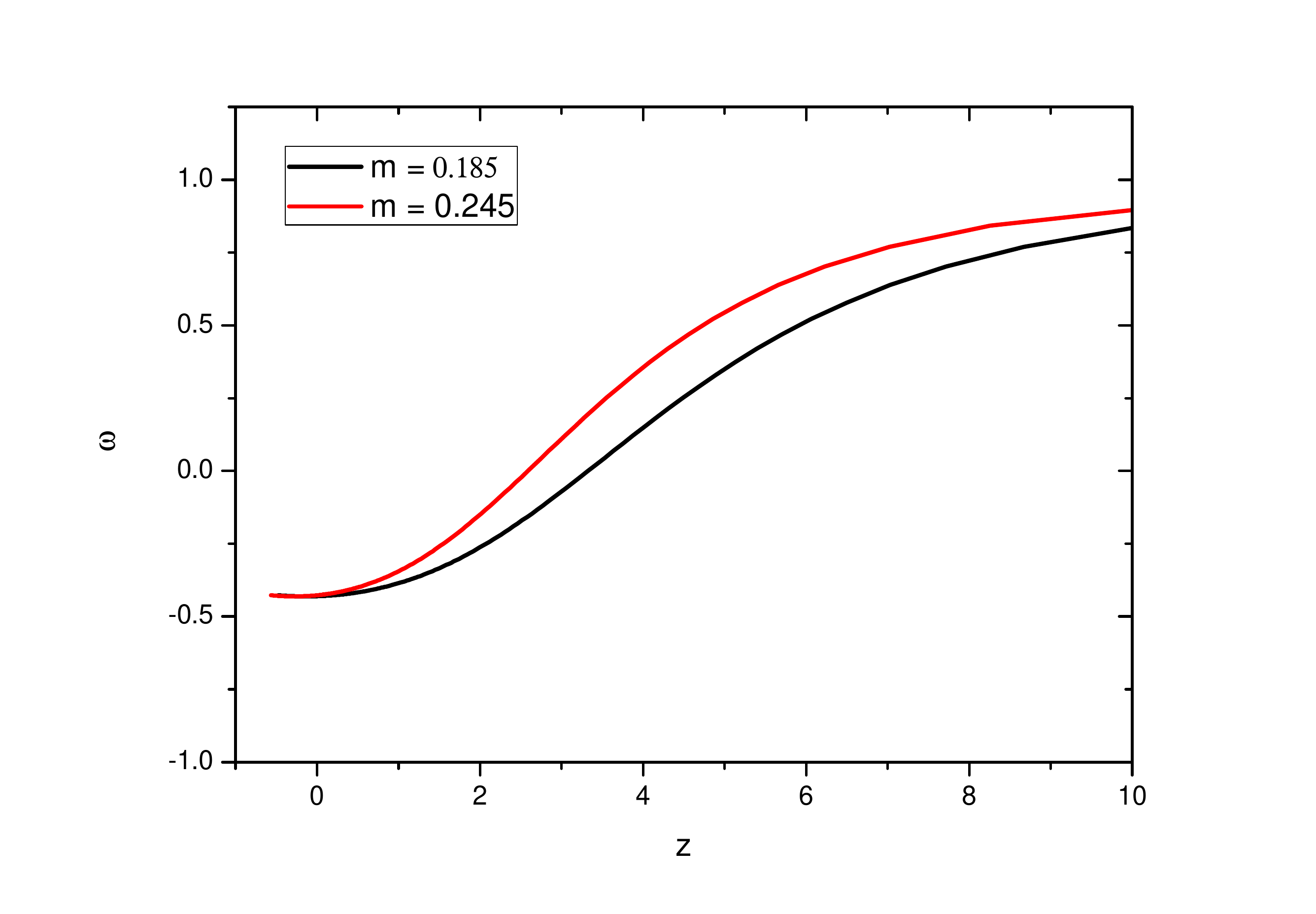}
\caption{Effective dark energy equation of state as a function of redshift for the choice of the functional $f(R,T)=\mu(R+T)$.}
    \end{center}
 \end{figure}

The energy density and the pressure of the model can be obtained from eqs. (11)-(13) and (15) as
\begin{equation}
\rho=\left(\frac{1}{\beta^2-\alpha^2}\right)\left[\frac{(\alpha-\beta)(2~lnt +1)^2+8\alpha(lnt)^2+2\beta}{4t^2(lnt)^2}\right],\label{eq:37}
\end{equation}
\begin{equation}
p=\left(\frac{1}{\beta^2-\alpha^2}\right)\left[\frac{(\beta-\alpha)(2~lnt +1)^2+8\beta(lnt)^2+2\alpha}{4t^2(lnt)^2}\right].\label{eq:38}
\end{equation}

Consequently, the effective dark energy equation of state (EoS) $\omega=\frac{p}{\rho}$ becomes
\begin{equation}
\omega= -1+\left[\frac{2(\alpha +\beta)\{1+4(lnt)^2\}}{(\alpha-\beta)(2~lnt +1)^2+8\alpha(lnt)^2+2\beta}\right].\label{eq:39}
\end{equation}

The effective dark energy equation of state evolves with time. However, the evolution of $\omega$ is governed by the two parameters $\alpha$ and $\beta$. It should be mentioned here that, different choice of these parameters leads to different model in the modified gravity. In the present work, we have considered two specific models of the modified gravity where the functional $f(R,T)$ is either $f(R,T)=R+2\lambda T$ or $f(R,T)=\mu (R+T)$. In the first case, $\alpha =8\pi+3\lambda$ and $\beta = \lambda$ whereas in the second choice $\alpha= \frac{16\pi+3\mu}{2\mu}$ and $\beta=0.5$. In both the cases, the effective dark energy equation of state parameter evolves from a positive value at the beginning to an asymptotic value $-1+\frac{2(\alpha+\beta)}{3\alpha-\beta}$ at late time.

The evolution of $\omega$, for the first case, is shown as a function of redshift $z$ in Figure 3. In the figure, we have shown different plots corresponding to different choices of the parameter $\lambda$. The choice of the parameter $m$ has already been constrained from the cosmic transit redshift. All the curves decrease from some positive value at an early epoch to $-\frac{\pi}{3\pi+\lambda}$ at late times of cosmic evolution. It is obvious that, with the decrease in the value of $\lambda$, the evolutionary trajectory lowers and also reaches to a low value at late time. For a given value of $\lambda$, $\omega$ evolves with higher values for early cosmic transit. One can note from the figure that, at any early epoch, the models appears to coincide, however, they follow different evolutionary path with the growth of cosmic time. For $\lambda =0$, the model reduces to the case of GR where $\omega$ decreases to $-\frac{1}{3}$. The models favour quintessence phase for the specific choices of the parameter $\lambda$. It is  interesting to note that, at late phase of cosmic time, the $f(R,T)$ gravity with a non zero value of $\lambda$, pulls the equation of state parameter downwards to make it closer to $-1$. But in the early phase, the behaviour of the equation of state parameter from $f(R,T)$ gravity is almost coincide with that from GR.

In Figure 4, we have shown the evolution of the effective dark energy equation of state $\omega$ for the second case with $f(R,T)=\mu (R+T)$. In order to get a simplified picture of the model, in the numerical calculation we have considered $\mu=1$. This particular choice leads to a beautiful correlation between the two models chosen for $f(R,T)$. In other words, for $\lambda=\frac{1}{2}$, the two models coincide. As is evident from the fig.4, $\omega$ for this model follows a similar trend as that of the previous one in fig.3. $\omega$ evolves from some positive value at any early epoch to an asymptotic value of $-\frac{2\pi}{6\pi+\mu}$ at late phase. For $\mu=1$, this asymptotic value becomes close to $-\frac{1}{3}$. One can compare the results for both the models from figs. 3 and 4. The dynamical evolution rate of the equation of state parameter is relatively more in the second case. This behaviour may be due to the fact that, in the second case we have rescaled the functional $f(R,T)$.

It is worth to mention here that, in our earlier work \cite{SKS16}, we have constructed some accelerating models at the back drop of an anisotropic BIII universe with the same choices of the functional for $f(R,T)$. One of the directional scale factor was considered from power law cosmology as $C=t^n$, $n \neq 1$ for all positive values of $n$. In that work, the mean scale factor was linear in cosmic time $t$ and the deceleration parameter was constant through out the cosmic evolution. Also, the equation of state parameter was found to be a constant quantity depending on the choice of the model parameters $n, \lambda$ and $\mu$. But it is quite interesting that, in the present work, we consider  the specific directional scale factor to be linear in cosmic time ($C=t$) and obtain a time varying deceleration parameter and a dynamically changing equation of state parameter. The time varying nature of the deceleration parameter enabled us to explore the cosmic transit from a decelerated phase of expansion to an accelerated one.

In terms of the metric potentials, the Ricci scalar $R$ for the BIII metric is expressed as
\begin{equation}
R=2\biggl(\frac{\ddot{A}}{A}+\frac{\ddot{B}}{B}+\frac{\ddot{C}}{C}+\frac{\dot{A}\dot{B}}{AB}+\frac{\dot{B}\dot{C}}{BC}+\frac{\dot{C}\dot{A}}{CA}-\frac{m^2}{A^2}\biggr),\label{eq:18}
\end{equation}
which for the present work becomes
\begin{equation}
R=\frac{(2lnt+1)(6lnt+1)-2}{2t^{2}(lnt)^{2}}.
\end{equation}
The scalar curvature $R$ goes to zero for large cosmic time. In terms of the model parameter $\alpha$ and $\beta$, the trace $T$ of the model is expressed as
\begin{equation}
T= \left(\frac{1}{\beta^2-\alpha^2}\right)\left[\frac{4(\alpha-\beta)(2~lnt +1)^2+8(\alpha-3\beta)(lnt)^2+2(\beta-3\alpha)}{4t^2(lnt)^2}\right].\label{eq:34}
\end{equation}
The trace of a given model can be obtained by using the respective values of the model parameters $\alpha$ and $\beta$. From the above expressions of the Ricci scalar and trace of energy-momentum tensor, we can construct the corresponding viable modified gravity model $f(R,T)$.

\section{Conclusion}
Dynamics of an anisotropic BIII universe is studied in the framework of a modified gravity, where the geometry part of the Einstein-Hilbert action contains some stuff of the matter field. This interesting aspect of the theory, called $f(R,T)$ gravity, have been investigated elsewhere. However, in the present work, we have explored the cosmic transit redshift as a constant of nature and constrained the parameters of the model to study the dynamics. In this context, we have considered two functional forms of $f(R,T)$: $f(R,T)=R+2\lambda T$ and $f(R,T)=\mu (R+T)$. The choice of these models is based upon the idea that they should, more or less, behave like GR. The first case reduces to GR for a very small value of $\lambda$ whereas the second case rescales the field equations in GR. Also these models inherently incorporate the time dependence of the gravitational constant and cosmological constant. Interestingly the two models overlap for some specific values of the model parameters. The field equations are integrated in a novel manner to get exact solutions. The model parameters that appear in the scale factors are constrained from the cosmic transit redshift. Interestingly, the modification of the Einstein-Hilbert action does not affect the scale factor and hence the deceleration parameter. However, the dynamics of the universe as calculated from the effective equation of state parameter is substantially affected  by the incorporation of the functional $f(R,T)$. Also, different choices of the functional $f(R,T)$ affect the dynamics. The equation of state parameter follows different trajectories for different values of the cosmic transit redshift. For a cosmic transit occurring in near past, the slope of $\omega$ becomes stiff compared to the case when the cosmic transit occurs at remote past.
\section*{Acknowledgment}

SKS would like to thank the BITS-Pilani, Hyderabad Campus, India for providing facility during an academic
visit where a part of this work was done.

\end{document}